%% file: main.tex
\documentclass[10pt,conference]{IEEEtran}
\IEEEoverridecommandlockouts
\usepackage{cite}
\usepackage{amsmath,amssymb,amsfonts}
\usepackage{algorithmic}
\usepackage{graphicx}
\usepackage{textcomp}
\usepackage{xcolor}
\usepackage{transparent}
\usepackage{diagbox}
\usepackage{multirow}
\usepackage{bm}
\usepackage{float} 
\usepackage{subcaption}
\usepackage{booktabs}
\usepackage{balance}
\usepackage[hidelinks]{hyperref}

\newcommand{\Junzhuo}[1]{{\color{magenta} #1}}	
\newcommand{\Ziwen}[1]{{\color{teal} #1}}

\def\BibTeX{{\rm B\kern-.05em{\sc i\kern-.025em b}\kern-.08em
    T\kern-.1667em\lower.7ex\hbox{E}\kern-.125emX}}
\begin{document}

\title{
\fontsize{20}{18}\selectfont \shortstack{
\hspace{-0.7cm}SetupKit: Efficient Multi-Corner \underline{Setup}/Hold Time \underline{Ch}aracterization\\
Using Bias-Enhanced \underline{I}nterpolation and Ac\underline{t}ive Learning}\\[-24pt]
}

\author{
\IEEEauthorblockN{
\href{https://orcid.org/0009-0009-6317-6373}{Junzhuo Zhou}$^{1}$,
\href{https://orcid.org/0009-0003-5797-7429}{Ziwen Wang}$^{2}$, 
\href{https://orcid.org/0009-0007-8115-1693}{Haoxuan Xia}$^{1}$, 
\href{https://orcid.org/0009-0006-5154-1845}{Yuxin Yan}$^{1}$,
Chengyu Zhu$^{3}$, 
\href{https://orcid.org/0000-0002-5208-482X}{Ting-Jung Lin}$^{2*}$,
\href{https://orcid.org/0000-0002-3177-8478}{Wei Xing}$^{4*}$,
\href{https://orcid.org/0000-0002-5266-3805}{Lei He}$^{1}$
}
\IEEEauthorblockA{
$^{1}$\textit{University of California, Los Angeles, United States} \\
$^{2}$\textit{Ningbo Institute of Digital Twin, Eastern Institute of Technology, Ningbo, China} \\
$^{3}$\textit{BTD Inc, Ningbo, China} $^{4}$\textit{The University of Sheffield, Sheffield, United Kingdom} \\
tlin@idt.eitech.edu.cn,
w.xing@sheffield.ac.uk\\[-24pt]}
}

\maketitle

\begin{abstract}

Accurate setup/hold time characterization is crucial for modern chip timing closure, but its reliance on potentially millions of SPICE simulations across diverse process-voltage-temperature (PVT) corners creates a major bottleneck, often lasting weeks or months. Existing methods suffer from slow search convergence and inefficient exploration, especially in the multi-corner setting. We introduce SetupKit, a novel framework designed to break this bottleneck using statistical intelligence, circuit analysis and active learning (AL). SetupKit integrates three key innovations: BEIRA, a bias-enhanced interpolation search derived from statistical error modeling to accelerate convergence by overcoming stagnation issues, initial search interval estimation by circuit analysis and AL strategy using Gaussian Process. This AL component intelligently learns PVT-timing correlations, actively guiding the expensive simulations to the most informative corners, thus minimizing redundancy in multi-corner characterization. Evaluated on industrial 22nm standard cells across 16 PVT corners, SetupKit demonstrates a significant 2.4$\times$ overall CPU time reduction (from 720 to 290 days on a single core) compared to standard practices, drastically cutting characterization time. SetupKit offers a principled, learning-based approach to library characterization, addressing a critical EDA challenge and paving the way for more intelligent simulation management.

\end{abstract}


\vspace{-0.3cm}
\let\thefootnote\relax
\footnotetext{\vspace{-0.3cm}$^{*}$Corresponding authors.}
\vspace{-0.1cm}
\section{Introduction}
\vspace{-0.1cm}
Accurate setup/hold time characterization is crucial for modern chip timing closure \cite{roethig2003library,phelps1991advanced}, but its reliance on potentially millions of SPICE simulations across diverse process-voltage-temperature (PVT) corners creates a major bottleneck, often lasting weeks or months. With sub-10nm designs requiring validation across 20-30 PVT corners, a mere 5ps inaccuracy can cause catastrophic timing failures in multi-GHz circuits \cite{Onaissi2011}.
For a typical library with 100 sequential cells, multiple pin combinations and varying conditions, constraint timing (\textit{e.g.}, setup/hold) consumes up to 80\% of total characterization time and thousands days on a single core \cite{Sharma:2015}.

Traditional optimization techniques for setup/hold characterization suffer from fundamental limitations. The bisection method, while robust, exhibits only linear convergence, requiring 15-20 simulations per point. Interpolation-based methods like Regula Falsi and quadratic interpolation offer theoretically faster convergence but frequently stagnate when confronted with steep metastability transitions common in advanced technologies, as they become trapped with test points near one interval endpoint in regions with high curvature.

Cadence's Liberate \cite{cadence2018liberate} implements an improved variant of Brent's method \cite{brent2013algorithms}, which attempts to combine bisection's reliability with interpolation's speed through dynamic switching between methods. However, our analysis reveals that even this approach only reduces simulation count by 10-15\% compared to pure bisection in real-world characterization scenarios, as the abrupt switching mechanism fails to address the root cause of interpolation's stagnation. Alternative analytical approaches such as those proposed by Srivastava et al. \cite{srivastava2008independent, srivastava2007rapid, srivastava2007interdependent} attempt to derive closed-form expressions for setup/hold time, but these methods struggle with modern transmission-gate-based register designs where the complexity of state transitions makes analytical modeling impractical and inaccurate.

Meanwhile, recent machine learning (ML) efforts to accelerate characterization have focused primarily on statistical sampling reduction. Naswali et al. \cite{Naswali:2019} leverage intra-table correlations to reduce Monte Carlo samples, while Ma et al. \cite{Ma:2024} exploit correlations between cells of different drive strengths. Zhou et al. \cite{zhou2025lvfgen} employ active learning (AL) to estimate statistical timing distributions. While these approaches reduce the number of required simulations, they fundamentally rely on ML predictions as final results, introducing unacceptable accuracy risks for signoff-quality libraries. Furthermore, none addresses the core computational burden of each individual setup/hold characterization point—the iterative search process itself—which dominates CPU time consumption.

We introduce SetupKit, a comprehensive framework that fundamentally reconceptualizes setup/hold characterization through a learning-driven, adaptive approach. Unlike previous methods that apply fixed search algorithms across all characterization points, SetupKit employs statistical intelligence to dynamically optimize the search process based on both circuit analysis and cross-corner learning. 
The novelties include:
\begin{itemize}
    \item \textbf{Bias-Enhanced Interpolation with Redundancy Adjustment (BEIRA)}: A novel root-finding algorithm that statistically models interpolation error and introduces an optimal bias term to prevent stagnation issues and achieve 1.3$\times$ faster convergence than bisection.
    
    \item \textbf{Circuit Analysis-Based Initial Interval Estimation}: A zero-simulation-cost technique that leverages logic effort analysis to predict setup/hold time bounds, reducing the initial search interval by up to 10$\times$ compared to conservative defaults in commercial tools.
    
    \item \textbf{Active Learning for Multi-Corner Characterization}: An uncertainty-driven framework that progressively learns the relationship between PVT parameters and setup/hold times, strategically selecting the most informative corners for simulation while predicting others. This component reduces the total simulation count by 40\% in multi-corner scenarios without compromising accuracy.
    
    \item \textbf{Comprehensive Experimental Validation}: Experiments on a production 22nm standard cell library across 16 PVT corners with full statistical variations demonstrate a substantial 2.4$\times$ overall CPU time reduction—shrinking characterization time from \textbf{720 days} to \textbf{290 days} on a single core for 4 million characterization points.
\end{itemize}

\vspace{-.3cm}
\section{Background and Preliminaries}\label{sec:prelim}
\vspace{-.1cm}

\begin{figure}[!t]
	\centering
		\includegraphics[width=7cm]{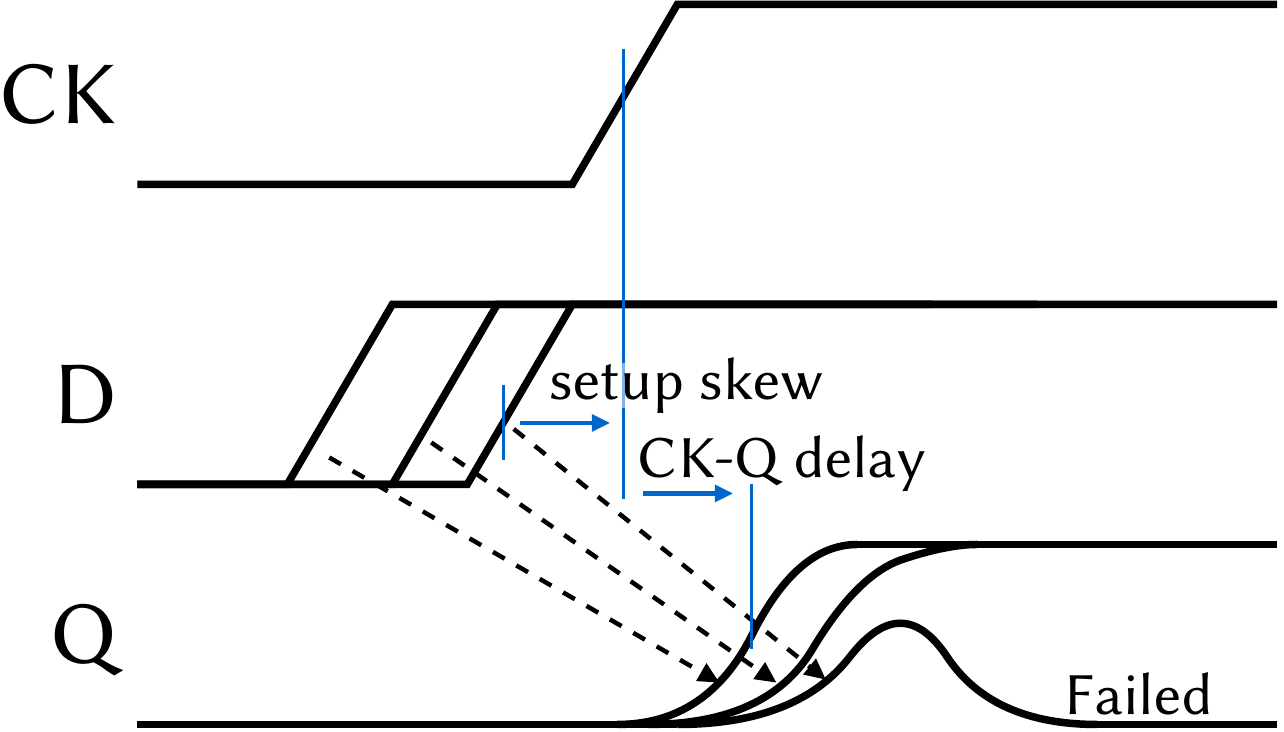}
        \vspace{-.1cm}
        \caption{Waveform behavior for different setup skews.}
        \vspace{-.4cm}
        \label{fig:wave}
\end{figure}
\begin{figure}[!t]
        \vspace{-.1cm}
	\centering
		\includegraphics[width=6.8cm]{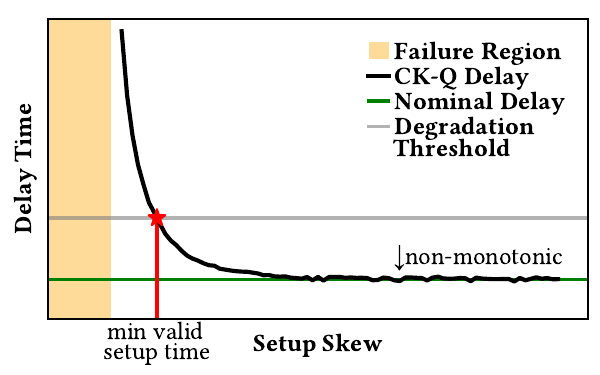}
        \vspace{-.3cm}
        \caption{CK-Q delay behavior for different setup skews.}
        \vspace{-.7cm}
        \label{fig:degradation}
\end{figure}
\subsection{Setup/Hold Time Characterization}
\vspace{-.1cm}
Setup and hold times are critical timing parameters that ensure reliable data capture in sequential circuits. The setup time defines the minimum duration data must be stable before a clock edge, while hold time specifies how long data must remain stable after the clock edge.

As illustrated in Fig.~\ref{fig:wave}, real circuits exhibit distinct behavior patterns as setup skew varies. When data arrives sufficiently early (large positive setup skew), the circuit operates normally with a consistent CK-Q delay. As the setup skew decreases toward a critical threshold, the CK-Q delay increases sharply due to metastability effects (Fig.~\ref{fig:degradation}). Beyond this threshold is the failure region where the circuit cannot properly capture the intended data value.

Industrial characterization tools define setup/hold times using two key criteria.
(1) Degradation criterion: The minimum setup/hold skew that ensures CK-Q delay remains below a specified threshold (typically 110\% of nominal delay).
(2) Pass-fail criterion: The minimum setup/hold skew that ensures correct data capture functionality.

The characterization process can be formalized as a root-finding problem:
Given a SPICE simulation function $f:X\to Y$ that maps setup skew space $X$ to CK-Q delay space $Y$, degradation threshold $y_0\in Y$, and target precision $\tau > 0$, where $f$ is monotonic around $y_0$, 
find $x_0 \in X$ such that:
\vspace{-.3cm}
\begin{equation}
   f(x_0) = y_0\\[-.3cm]
\end{equation}
The solution must be approximated within a specified tolerance, finding $\hat{x}_0$ such that $|\hat{x}_0 - x_0| < \tau$.
This characterization has to be performed across all cells in a standard cell library, for all possible combinations of input slew, output load, PVT corners, and process variations, resulting in millions of SPICE simulations.

\vspace{-.2cm}
\subsection{Bracketing Interval Search Methods}\label{sec:search}
\vspace{-.1cm}
\begin{figure}[!t]
	\centering
		\includegraphics[width=7.5cm]{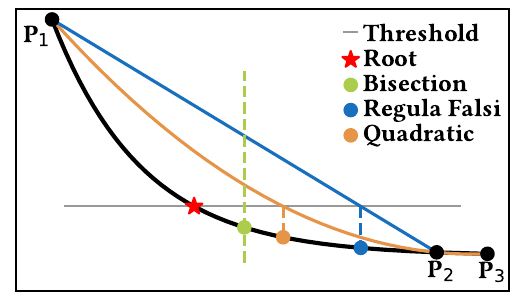}
        \vspace{-.1cm}
        \caption{Test points obtained from different search methods. Given root's Y location threshold; two interval endpoints P$_1$ and P$_2$;  outside adjacent point P$_3$.}
        \vspace{-.5cm}
        \label{fig:search}
\end{figure}

An efficient search method is significant since each iteration step of searching means a costly transient SPICE simulation.

\noindent\textbf{Bisection Method:} The most robust approach that iteratively halves the search interval, achieving linear convergence with time complexity $O(\log(1/\tau))$. Its simplicity and reliability have made it the standard in industrial tools, as it works for both degradation and pass-fail criteria.


\noindent\textbf{Interpolation Methods:} Techniques like Regula Falsi (linear interpolation) and quadratic interpolation select test points based on function values at known points (Fig.~\ref{fig:search}). While theoretically faster than bisection under ideal conditions, they suffer from a critical weakness: when the root is close to an interval endpoint or when the function has high curvature, these methods can get trapped selecting test points repeatedly near the same endpoint, degrading to sublinear convergence.


\noindent\textbf{Brent's Method \cite{brent2013algorithms}:} This hybrid approach, implemented in tools like Cadence Liberate, attempts to combine the reliability of bisection with the efficiency of interpolation. It dynamically switches between interpolation and bisection based on the convergence behavior, preventing worst-case scenarios but still exhibiting suboptimal performance in many practical cases.

However, there are two issues for today's characterization implementations.
Firstly, the initial search interval is typically fixed as a huge range with a large search step to ensure the interval is closure.
Secondly, the convergence of search methods can be further improved.
\vspace{-.2cm}
\section{Methodologies: BEIRA and Self-Adaptive}\label{sec:beira}
\vspace{-.1cm}
Traditional root-finding methods face a critical trade-off: bisection guarantees linear convergence but ignores function values, while interpolation methods potentially converge faster but can stagnate near interval boundaries. We introduce Bias-Enhanced Interpolation with Redundancy Adjustment (BEIRA), a novel method that overcomes these limitations by statistically modeling interpolation error and strategically biasing test points.


\begin{figure}[!t]
	\centering
		\includegraphics[width=8cm]{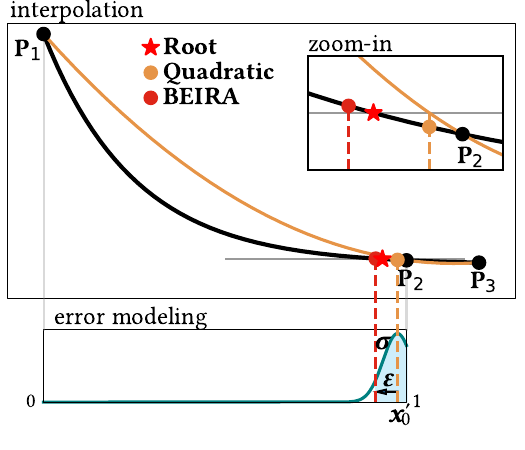}
        \vspace{-.6cm}
        \caption{Above: quadratic interpolation. Bottom: finding the bias $\varepsilon$ from the distribution of $x_0$, assuming the estimation error is under Gaussian distribution.}\label{fig:redundancy}
        \vspace{-.7cm}
\end{figure}
\subsection{Statistical Modeling of Interpolation Error}
\vspace{-.1cm}

The key insight of BEIRA is treating interpolation uncertainty as a probabilistic problem. In Fig.~\ref{fig:redundancy}, we normalize the search interval so that endpoints P$_1$ and P$_2$ have coordinates $x_1=0$ and $x_2=1$. When quadratic interpolation suggests a test point $x_0'$ (especially one near an endpoint), it may not be optimal for efficient interval reduction.

BEIRA treats the true root location as a random variable:
\vspace{-.2cm}
\begin{equation}
\hat{x}_0 \sim \mathcal{N}(x_0',\sigma^2),\text{ subject to } \hat{x}_0\in[0,1].\\[-.25cm]
\end{equation}
Here $x_0'$ is current estimate from interpolation, and $\sigma$ represents uncertainty. By adding a bias term $\varepsilon$ to create a test point at $x_0'+\varepsilon$, we can strategically optimize the search process.
\vspace{-.2cm}
\subsection{Finding the Optimal Bias}
\vspace{-.1cm}
When we test at position $x_0'+\varepsilon$ (where $\varepsilon \in [-x_0',1-x_0']$), two outcomes are possible: (1) the true root lies in $[0,x_0'+\varepsilon]$, giving new interval length $x_0'+\varepsilon$; (2) the true root lies in $[x_0'+\varepsilon,1]$, giving new interval length $1-x_0'-\varepsilon$.
The probability of each case depends on our uncertainty model:
\vspace{-.2cm}
\begin{equation}
\begin{aligned}
P_1 &= P(\hat{x}_0 \leq x_0'+\varepsilon) - P(\hat{x}_0 < 0) \\[-.1cm]
&= \Phi\left(\frac{\varepsilon}{\sigma}\right) - \Phi\left(-\frac{x_0'}{\sigma}\right)
\end{aligned}
\end{equation}
\vspace{-.3cm}
\begin{equation}
\begin{aligned}
P_2 &= P(\hat{x}_0 \leq 1) - P(\hat{x}_0 \leq x_0'+\varepsilon) \\[-.1cm]
&= \Phi\left(\frac{1-x_0'}{\sigma}\right) - \Phi\left(\frac{\varepsilon}{\sigma}\right),\\[-.2cm]
\end{aligned}
\end{equation}
where $\Phi$ is the standard normal cumulative distribution function.
The expected new interval length is:
\vspace{-.2cm}
\begin{equation}
E[L] = P_1 \cdot (x_0'+\varepsilon) + P_2 \cdot (1-x_0'-\varepsilon).\\[-.3cm]
\end{equation}
We want to minimize this expected length:
\vspace{-.2cm}
\begin{equation}
\min_{\varepsilon} E[L] \quad \text{subject to } \varepsilon \in [-x_0',1-x_0'],\\[-0.2cm]
\end{equation}
which is equivalent to:
\vspace{-.2cm}
\begin{equation}
\begin{aligned}
\frac{dE[L]}{d\varepsilon} &= P_1 + (x_0'+\varepsilon) \frac{dP_1}{d\varepsilon} - P_2 + (1-x_0'-\varepsilon) \frac{dP_2}{d\varepsilon} = 0 \\[-.3cm]
\end{aligned}
\end{equation}
Substituting the derivatives and simplifying:
\vspace{-.2cm}
\begin{equation}\label{eq:derivative}
\begin{aligned}
\frac{dE[L]}{d\varepsilon} &= 2\Phi\left(\frac{\varepsilon}{\sigma}\right) - \Phi\left(\frac{1-x_0'}{\sigma}\right) - \Phi\left(-\frac{x_0'}{\sigma}\right) \\[-.1cm]
&+ \frac{1}{\sigma}\phi\left(\frac{\varepsilon}{\sigma}\right)[2(x_0' + \varepsilon) - 1] = 0,
\end{aligned}
\end{equation}
where $\phi$ is the standard normal probability density function.
For practical implementation, we use this approximation when $\sigma$ is small:
\vspace{-0.4cm}
\begin{equation}\label{eq:appro}
\varepsilon^* \approx
\begin{cases}
-\sigma\sqrt{2\ln\left(\frac{2x_0'-1}{\sigma\sqrt{2\pi}}\right)}, & \text{if } x_0' > \frac{1}{2}\\[2mm]
\sigma\sqrt{2\ln\left(\frac{1-2x_0'}{\sigma\sqrt{2\pi}}\right)}, & \text{if } x_0' < \frac{1}{2}\\[2mm]
0, & \text{if } x_0' = \frac{1}{2} \\[-.2cm]
\end{cases}
\end{equation}

This bias strategically positions test points to maximize expected convergence, unlike methods that simply trust interpolation estimates.
\vspace{-0.2cm}
\subsection{Self-Adaptive Mechanism}
\vspace{-0.1cm}
\begin{figure}[!t]
	\centering
    \vspace{-0.1cm}
	\includegraphics[width=8cm]{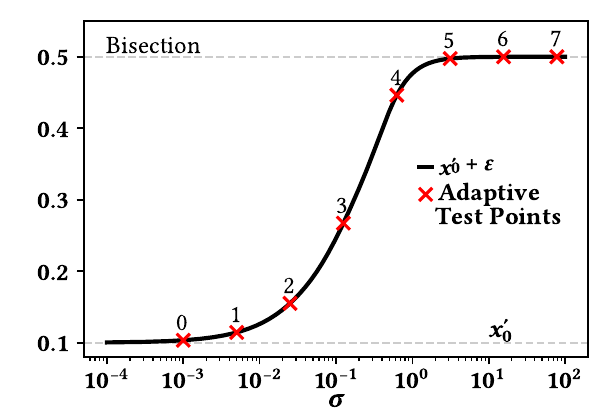}
    \vspace{-0.3cm}
	\caption{As uncertainty ($\sigma$) increases from 0.001 to 100, BEIRA smoothly transitions from interpolation-based test points to bisection-like behavior ($x_0'=0.1$, $\beta=5$, for $n = 0,...,7$).}\label{fig:adaptive}
    \vspace{-0.5cm}
\end{figure}

Instead of abruptly switching between methods like Brent's algorithm, BEIRA gradually adapts when it detects potential stagnation. When test points repeatedly fall on the same side of the interval, BEIRA increases its uncertainty parameter:
\vspace{-0.2cm}
\begin{equation}
\sigma = \sigma_0 \cdot \beta^n\\[-2.5mm]
\end{equation}
where $\sigma_0$ is the initial uncertainty (typically 0.001), $\beta$ is the growth factor (typically 5), and $n$ is the number of consecutive iterations stuck on one side.

As shown in Fig.~\ref{fig:adaptive}, increasing $\sigma$ shifts test points toward the interval midpoint (0.5 in normalized coordinates). This creates a smooth transition from aggressive interpolation-based searching to conservative bisection-like behavior without arbitrary switching rules.
This adaptive approach effectively handles challenging cases where standard interpolation methods struggle, such as when the root lies near an interval boundary, while maintaining efficiency when conditions are favorable.
\vspace{-0.2cm}
\section{Initial Search Interval Estimation}\label{sec:estimation}
\vspace{-0.1cm}
The initial search interval can be formularized as an initial test location $l_0$ and the initial step $s_0$, where $l_0$ will be assigned as the first test point, followed by $l_0-2^ns_0$ or $l_0+2^ns_0$ toward the other side of the threshold. Here, $n=0,1,2,...$ represents the attempt count, increasing until the opposite interval endpoint is obtained.


As shown in Fig.~\ref{fig:degradation}, the relationship between setup skew and CK-Q delay is not smooth and can even be non-monotonic for large setup skews, making interpolation infeasible for obtaining initial search intervals. Current characterization implementations typically use a fixed, overly conservative initial interval with large $l_0$ and $s_0$ values to ensure interval closure, unnecessarily increasing the number of search iterations.
SetupKit addresses this inefficiency through two complementary approaches: (1) a circuit analysis-based method requiring zero simulations for general characterization scenarios, and (2) an AL framework that leverages cross-corner correlations for multi-corner scenarios. Both strategies drastically reduce the number of iterations needed to establish a bracketing interval containing the true setup/hold time.

\vspace{-0.2cm}
\subsection{Initial Search Interval Estimation By Circuit Analysis}\label{sec:circuit}
\vspace{-0.1cm}
\begin{figure}[!t]
	\centering
    \vspace{-1.2cm}
    \begin{subfigure}{1pt}
    \transparent{0}
    \caption{}
    \label{fig:latch}
    \caption{}
    \label{fig:dff}
    \end{subfigure}
		\includegraphics[width=9cm]{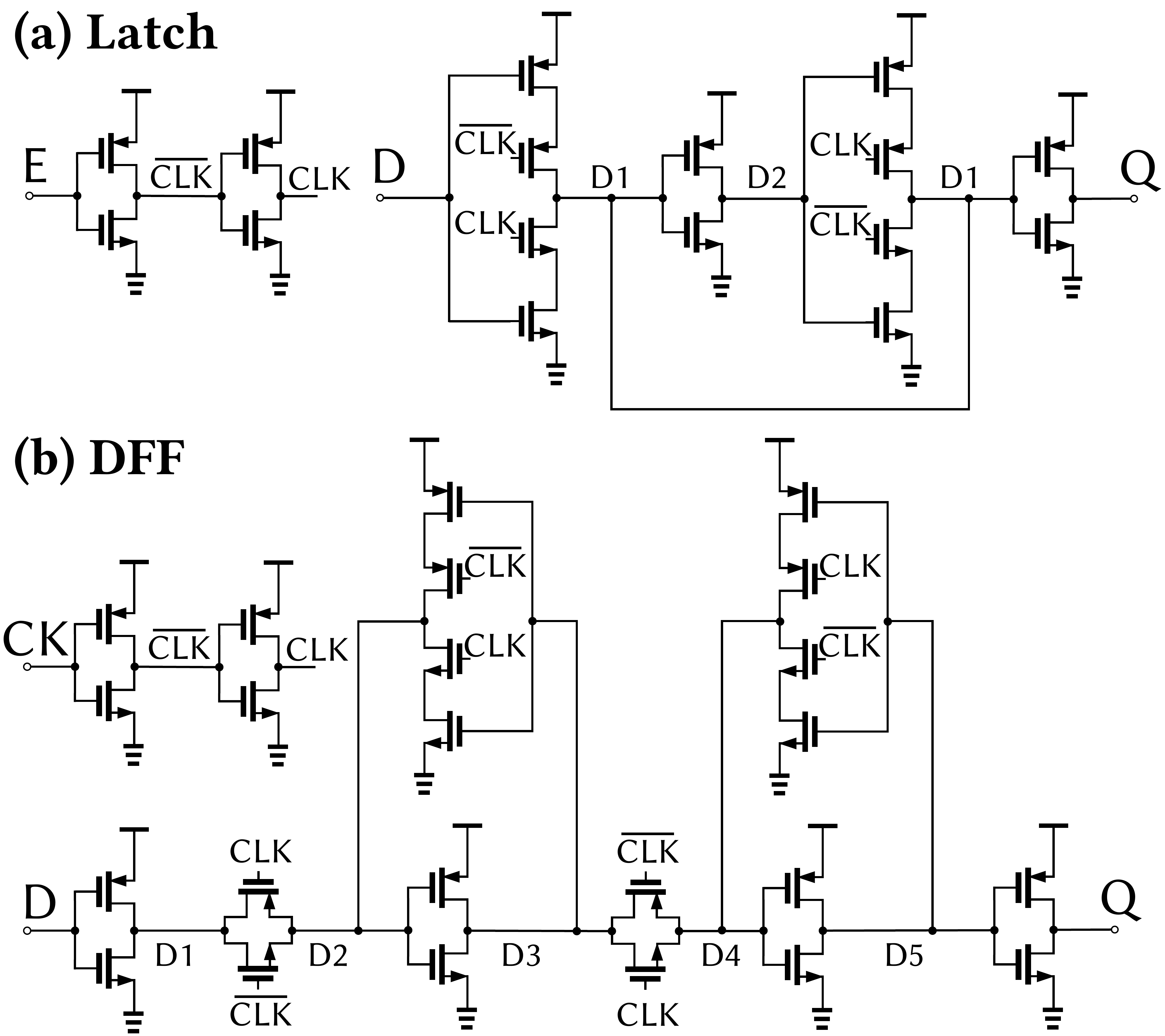}
        \vspace{-0.4cm}
        \caption{Typical schematic of (a) latch and (b) DFF.}
        \vspace{-0.6cm}
        \label{fig:latch_dff}
\end{figure}
We propose a formula-based method that analyzes logic effort to estimate the initial search interval for the setup/hold time. Unlike Liberate's path-difference method that directly estimates setup/hold time, SetupKit adopts estimation only for the initial search interval. The simulation of nominal CK-Q delay is the first and unavoidable step for all setup/hold characterizations; SetupKit utilizes this delay time and the relation between delay and setup/hold to estimate initial search intervals.
We use the logic effort analysis as follows:
\vspace{-0.3cm}
\begin{equation}
\begin{aligned}
D = g \cdot h + p \cdot \gamma \\[-2.5mm]
\end{aligned}
\end{equation}
where $g$ is the logic effort, $h$ is the electrical effort (fan-out), and $p$ is the parasitic delay. We assume parasitic parameter $\gamma$ is 1, and all the gates in Fig.~\ref{fig:latch_dff} have the pull-up equal to pull-down (width of NMOS:PMOS is 1:2). The NMOS:PMOS width of transmission gates is 1:1.
We denote assumptions as:
\vspace{-0.35cm}
\begin{equation}
\begin{aligned}
g_{\text{TG}} = 2,\quad g_{\text{INV}} = 1,\quad p_{\text{TG}} = 2,\quad p_{\text{INV}} = 1,\quad h_i = 1 \\[-2.5mm]
\end{aligned}
\end{equation}

\textbf{Latch:}
Fig.~\ref{fig:latch} shows the typical schematic of a C$^2$MOS latch. We can use logic effort to calculate the nominal E-D delay and setup/hold time (in the unit of inverter delay).

\begin{itemize}
  \item Nominal E-D delay can be approximated by path:\\
$\texttt{E}\rightarrow\overline{\texttt{CLK}}\rightarrow\texttt{D}\rightarrow\texttt{D1}\rightarrow\texttt{D2}\rightarrow\texttt{D3}\rightarrow\texttt{Q}$\\
The approximated delay is $t_{\text{E-Q}} \approx 28$ units.
  \item Setup time using the path:
$\texttt{D}\rightarrow\texttt{D1}\rightarrow\texttt{D2}$\\
The approximated setup time is $t_{\text{setup}}\approx 12$ units.
  \item Hold time using the path:
$\texttt{D}\rightarrow\texttt{D1}$\\
The approximated hold time is $t_{\text{hold}} \approx 10$ units.
\end{itemize}

For latch's characterization, we assign the initial search interval for setup as
$l_0=s_0=0.4\cdot t_{\text{E-Q}}$, and $l_0=s_0=0.35\cdot t_{\text{E-Q}}$ for the hold time.

\textbf{DFF:}
Fig.~\ref{fig:dff} shows the typical schematic of a C$^2$MOS \& transmission gate DFF. We use logic effort to calculate the nominal CK-Q delay and setup/hold time (in the unit of inverter delay).

\begin{itemize}
  \item Nominal CK-Q delay can be approximated by path:\\
$\texttt{CK}\rightarrow\overline{\texttt{CLK}}\rightarrow\texttt{CLK}\rightarrow\texttt{D3}\rightarrow\texttt{D4}\rightarrow\texttt{D5}\rightarrow\texttt{Q}$\\
The approximated delay is $t_{\text{CK-Q}} \approx 10$ units.
  \item Setup time using the path: $\texttt{D}\rightarrow\texttt{D1}\rightarrow\texttt{D2}\rightarrow\texttt{D3}$

The approximated setup time is $t_{\text{setup}}\approx 7$ units.
  \item Hold time using the path: $\texttt{D}\rightarrow\texttt{D1}\rightarrow\texttt{D2}$

The approximated hold time is $t_{\text{hold}} \approx 3.33$ units.

\end{itemize}

For DFF's characterization, we can assign the initial search interval of setup time as
$l_0=s_0=0.7\cdot t_{\text{CK-Q}}$;
and $l_0=s_0=0.33\cdot t_{\text{CK-Q}}$ for the hold time.

In the previous analysis, we assume ideal signal transitions at gate inputs. However, in real circuits, the input signal slew can significantly affect the delay of logic gates. Especially when both the clock and pin D are switching, but with deviated slew rates, the nominal CK-Q delay can become a negative value. This phenomenon becomes more pronounced in deep submicron technologies where parasitic capacitance and resistance are non-negligible. We address this by adding a minimal constraint for $s_0$ to avoid negative search. With this circuit analysis-based initial interval estimation, SetupKit can rapidly find a narrow search interval without additional SPICE simulations, significantly improving practical efficiency.



\vspace{-2mm}
\subsection{Initial Search Interval Estimation By Active Learning (AL)}
\vspace{-1mm}
For characterization with sufficient PVT samples, regression becomes feasible to estimate the initial interval $\boldsymbol{y}_n$ of a new sample with its PVT information $\boldsymbol{x}_n$, by exploiting the correlations between previously simulated setup times $\mathcal{Y}$ and corresponding PVT information $\mathcal{X}$.

As depicted in Fig.~\ref{fig:AL}, our proposed AL component involves a high-level iteration framework, with $N$ total samples, hyperparameter batch size $M$ and maximum iteration number $k_{max}$. For generality, both global process corners and local process variations are taken into consideration. The AL framework progressively selects the samples that are expected to have large initial intervals and higher impacts on the regression model. The framework consists of three main steps for each AL iteration: sample selection, simulations for the setup time search, and regression to estimate initial intervals. These steps are detailed below:

\begin{figure}[!t]
        \vspace{-2mm}
	\centering
        \includegraphics[width=9cm]{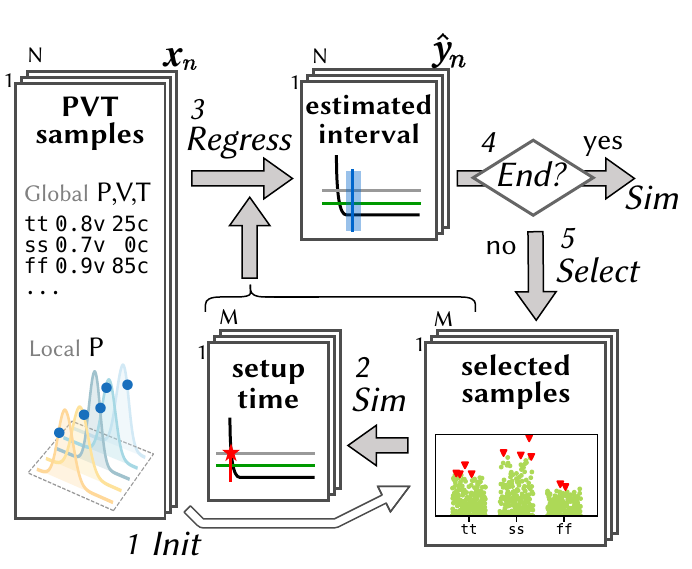}
        \vspace{-7.5mm}
        \caption{Overview of active learning regression to estimate initial search interval.}\label{fig:AL}
        \vspace{-7mm}
\end{figure}

\input{table/corners}

\subsubsection{Initial Sample Selection (set $k=0$)}\label{sec:init-select}
\begin{itemize}
\item Given PVT samples $\mathcal{X}=\{\boldsymbol{x}_n|\ n=1,2,...,N\}$.
\item Uniformly select $M$ samples to form the initial sample set $\mathcal{X}_0=\{\boldsymbol{x}_n|\ n=\lfloor\frac{N}{M}\rfloor,\lfloor\frac{2N}{M}\rfloor,...,\lfloor\frac{MN}{M}\rfloor\}$. 
\item For each sample, estimate initial search interval by Section \ref{sec:circuit}'s circuit analysis. To step \textit{2)}.
\end{itemize}

\subsubsection{Simulations}\label{sec:sim}
\begin{itemize}
\item Search the setup time for samples $\mathcal{X}_k$ under the estimated initial interval with SPICE simulations. Use BEIRA or bisection to obtain setup time results $\mathcal{Y}_k$. To step \textit{3)}.
\end{itemize}
\subsubsection{Regression}\label{sec:regress}
\begin{itemize}
\item Train a Gaussian Process (GP) with simulated sample set $\{\mathcal{X}_0,...,\mathcal{X}_k\}$ and $\{\mathcal{Y}_0,...,\mathcal{Y}_k\}$.
\item Predict the setup time of not simulated samples in $\mathcal{X}$ with GP regression, obtain each sample's setup time expectation and variance
$\hat{\mathcal{Y}}=\{\hat{\boldsymbol{y}}_n|\  \hat{\boldsymbol{y}}_n\sim\mathcal{N}(\mu_n,v_n^2)\}$.
\item Estimate initial search interval for each sample $x_0$, 
use setup time expectation $\mu_n$ as initial test location, use uncertainty $v_n$ as initial step. To step \textit{4)}.
\end{itemize}

\subsubsection{Termination Criterion Check}\label{sec:end}
\begin{itemize}
\item If $k > k_{max}$,
terminate and simulate for all rest samples $\mathcal{X}$ with estimated initial intervals, get $\mathcal{Y}$, same to step \textit{2)}.
\item Else, proceed to step \textit{5)}.
\end{itemize}

\subsubsection{Sample Selection (set $k=k+1$)}\label{sec:select}
\begin{itemize}
\item For each PVT$_i$, denote $m_i=MV_i/V$, where $V=\sum v_n$ is the sum of all samples, $V_i$ is PVT$_i$'s sum uncertainty.
\item Select top-$m_i$ of sample uncertainty $v_n$ for each PVT$_i$.
\item Collect each PVT's selected samples into $\mathcal{X}_k$. To step \textit{2)}.
\end{itemize}

\input{fig/single}

For step \textit{5)}, a higher uncertainty $v_n$ is usually more likely to improve the regression. However, global maximum top-$M$ selection is inefficient in practice,
and our local maximum of top-$m_i$ selection within each corner is much more informative.

The AL framework is inherently parallelizable for two key reasons: (1) each iteration involves $M$ setup time searches that can be executed in parallel, and (2) the AL typically requires characterizing only a small fraction of the total samples, leaving the remaining $N-k_{max}\cdot M$ samples to be searched in parallel with their estimated initial intervals.

By strategically guiding expensive simulations to the most informative corners, SetupKit's AL approach minimizes redundancy in multi-corner characterization while maintaining SPICE-level accuracy for all characterized values. 
Validated in Section \ref{sec:exp-cross-pvt}, the AL approach reduces the average iteration count from 14.2 to 6.7 ($M = 200$, $k_{max} = 5$), representing a 53\% reduction in simulation cost with negligible overhead.

\vspace{-2mm}
\section{Experiments}\label{sec:exp}
\vspace{-1.5mm}
The experiments are carried out using standard TSMC 22 nm cells in 16 global PVT corners listed in Table~\ref{tab:corner}, with all local variations enabled.
All transient simulations are performed using HSPICE on Linux machines equipped with Intel Xeon 6348 CPUs. The proposed AL strategy is guided by GP implementations in SMT\footnote{\url{ https://github.com/SMTorg/smt}}.

We compare five search methods: BEIRA, bisection, quadratic interpolation, Brent's method, and the improved Brent's method implemented in Liberate.
BEIRA adopts typical parameters $\sigma_0=0.001$ and $\beta=5$.
All methods terminate upon reaching the target precision $\tau=0.01$ ps (simulator's precision).
The setup times of {\verb|DFCNQD1|} and {\verb|LHQD1|} cells are characterized as representative register and latch cases. The proposed method is applicable to other sequential timing metrics such as hold, removal, and recovery times.

\vspace{-2mm}
\subsection{General Characterization on Latch and DFF}
\vspace{-1mm}

\input{fig/iteration}
This experiment validates SetupKit for a single setup time characterization of DFF and latch, under the {\verb|tt0p8v25c|} PVT corner with nominal local process variations.
SetupKit estimates the initial search interval through circuit analysis for BEIRA method and other methods, whereas Liberate relies on a fixed initial interval.

Figs.~\ref{fig:dff-delay}, \ref{fig:latch-delay} illustrate the setup skew vs. delay behavior for the DFF and latch, respectively.
In both cases, the setup time is extremely close to the failure boundary. Notably, for the latch, the failure region even overlaps with valid delay points, frequently triggering a fallback to bisection due to ambiguity in the pass/fail criterion.
The steep curvature in the delay transition also increases the risk of failed convergence for interpolation-based methods.

Figs.~\ref{fig:dff-convergence}, \ref{fig:latch-convergence} present the convergence traces of all methods.
Typically, three simulations are sufficient to identify a valid search interval. Compared to fixed intervals, circuit analysis–based estimation provides $\sim 10\times$ higher initial precision.

During the iterations, all interpolation-based methods initially experience stagnation before rapidly shrinking the search interval, whereas bisection converges linearly throughout. BEIRA consistently converges in fewer than 10 steps, significantly outperforming other methods by escaping stagnation earliest and achieving the fastest convergence. In particular, the Liberate methods, which represent state-of-the-art commercial tools, require nearly twice the steps to reach same precision.

These results validate the effectiveness of circuit analysis–based interval initialization and the robustness of BEIRA, particularly in scenarios involving high-curvature delay transitions near the failure boundary. The proposed method is thus well-suited for general setup and hold time characterization.

\vspace{-2mm}
\subsection{Characterization Across PVT Corners \& Process Variations}\label{sec:exp-cross-pvt}
\vspace{-1mm}

This experiment assesses SetupKit by characterizing the {\verb|DFCNQD1|} cell ($168$-dimension) across PVT corners and process variation samples.
We generate $10k$ $168$-dimensional local process variation samples for each global PVT corner by Sobol's Quasi-Monte Carlo \cite{sobo:1967}, resulting in a total of $N=160k$ $171$-dimensional PVT samples. {\verb|TT|}, {\verb|FF|}, {\verb|SS|} are quantized into $0$, $1$, $-1$.
SetupKit searches setup time with BEIRA and estimates the initial search intervals by AL with batch size $M=200$ and the maximum iteration number $k_{max}=5$.

Fig.~\ref{fig:iter} presents the status of each step during AL iteration, involving sample selection, simulations, and regression.
Begin with the uniform selection, Fig.~\ref{fig:iter:selection} evidences our selection strategy successfully identifies samples with the largest initial search step (uncertainty) among each corner, with a balanced proportion.
The regression uncertainty decreases along the iterations.
Fig.~\ref{fig:iter:simulation} shows all precision traces in BEIRA search, in which we can observe the traces are rapidly compressed to the left side with fewer iterations.
Those traces are statistically steeper than bisection's reference line, demonstrating the faster statistical convergence of BEIRA.
The Pred. vs Real scatters in Fig.~\ref{fig:iter:regression} indicate the precision of initial interval estimation improves rapidly, verifying the efficiency of SetupKit's AL strategy.
The real setup time is only for evaluation as those plots will not exist in real deployment.

We use the predicted initial intervals at iteration 4 to acquire the setup time for the rest $159k$ samples with BEIRA. This results in an average simulation count of $6.67$ for each sample. The overall average simulation counts for the $160k$ samples is $6.69$, showing a significant improvement compared to $14.16$ in the initial iteration. The results indicate that SetupKit can precisely estimate the initial interval and intelligently learn PVT-timing correlations, actively guiding the expensive simulations to the most informative corners, thus minimizing redundancy in multi-corner characterization.

\begin{figure}[!t]
    \vspace{-4mm}
	\centering
		\includegraphics[width=9.5cm]{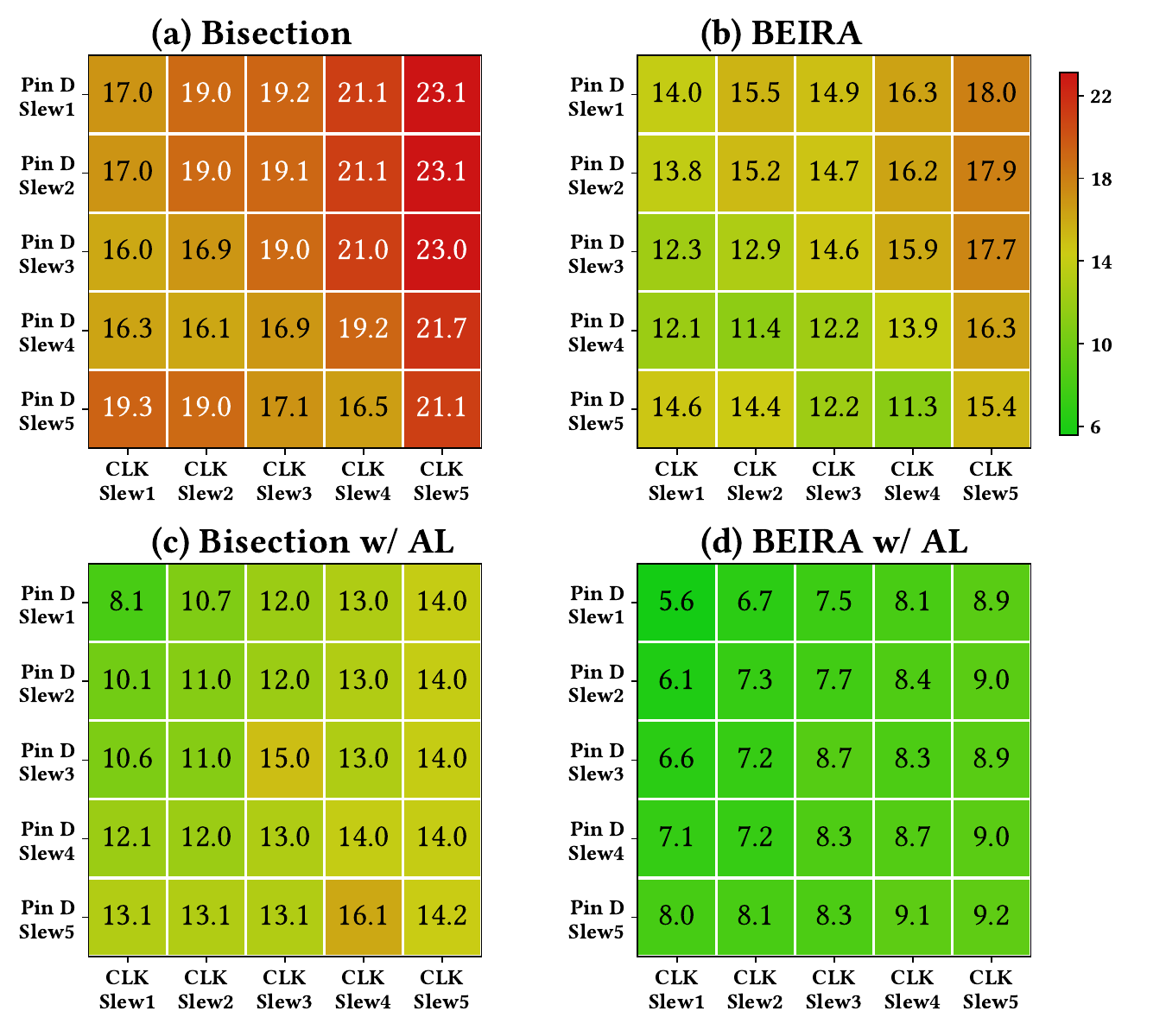}
        \caption{Comparison of final local iteration (simulation) count. Each value is the average result of $16\text{ PVT}\times10k\text{ MC}=160k$ samples.}\label{fig:compare}
        \vspace{-5mm}
\end{figure}

\subsection{Overall Runtime Comparison}

Here we present a comprehensive analysis of SetupKit's advantages. First, we extend the experiments to cover a $5\times5$ slew-slew table for the {\verb|DFCNQD1|} cell. Second, we assume two iteration methods (Bisection and BEIRA) and two initial interval settings (fixed and AL-prediction), to evaluate the performance of their combinations. 

Fig.~\ref{fig:compare} compares the simulation number required in each case. The values are the average results of $16\text{ PVT}\times10k\text{ MC}=160k$ samples. Note that the result of (D Slew1, CLK Slew2) shown in Fig.~\ref{fig:compare} (d) is obtained from the previous experiment in Section \ref{sec:exp-cross-pvt}. Those tables indicate that AL saves about 7 simulations and BEIRA saves another 4, showing the improvement of SetupKit is statistically significant.

For a fair comparison, Table~\ref{tab:compare} summarizes the breakdown of CPU time, including the runtime of AL and BEIRA's search.
Although the BEIRA is more complex and slower than bisection, the overheads are negligible compared to the simulation costs. The same principle is applicable for amortized AL CPU time.
SetupKit reduces the total CPU time from bisection's 15.5s per characterization to 6.4s, achieving a 2.4$\times$ speedup.
Considering the result is obtained by averaging $25\text{ entry}\times16\text{ PVT}\times10k\text{ MC}=4M$ samples, SetupKit reduces the total CPU time from 720 days to 290 days.
SetupKit can be accelerated easily by parallel computing because the dependency between algorithm components is negligible.
\input{table/compare}

\section{Conclusions}\label{sec:conclusion}

We have presented SetupKit, a learning-driven framework for multi-corner setup/hold time characterization. The proposed BEIRA algorithm statistically balances interpolation and bisection through optimal bias. With precise interval estimation from circuit analysis and AL for PVT-aware initialization, SetupKit reduces the overall runtime for $4M$-samples from 720 to 290 days (2.4× speedup) on a single core. 




Looking forward, several challenges remain to be addressed. The statistical modeling in BEIRA could be further improved through hardware-aware optimization of the bias computation, potentially using table-lookup approaches to reduce overhead. For multi-corner scenarios, the AL framework could be extended to incorporate variation-aware methods that better handle extreme corners and rare events. Future work will focus on integrating 
SetupKit with commercial EDA flows and extending our learning-driven paradigm to other characterization tasks, including removal/recovery timing and min-pulse-width, potentially revolutionizing how the industry approaches the entire library characterization process.

\newpage
\bibliographystyle{unsrt}
\balance
\bibliography{reference}
\end{document}

%% file: table/corners.tex
\begin{table*}[]
    \centering
    \vspace{-0.4cm}
    \caption{PVT Corners in Experiment$^\ast$}
    \vspace{-5pt}
    \label{tab:corner}
    \setlength{\tabcolsep}{4pt}
    \begin{tabular}{|l|cccccccccccccccc|}
        \hline
        \\[-7pt]
        \diagbox[width=2cm, height=0.55cm]{}{\textbf{Corner}}
         & \textbf{1} & \textbf{2} & \textbf{3} & \textbf{4} & \textbf{5} & \textbf{6} & \textbf{7} & \textbf{8} & \textbf{9} & \textbf{10} & \textbf{11} & \textbf{12} &\textbf{13} &\textbf{14}&\textbf{15}&\textbf{16}  \\ \hline
        \\[-7pt]
        \textbf{P}rocess & \texttt{TT} & \texttt{TT} & \texttt{TT} & \texttt{TT} & \texttt{FF} & \texttt{FF} &\texttt{FF} & \texttt{FF} & \texttt{FF} & \texttt{FF} & \texttt{SS} & \texttt{SS} & \texttt{SS} & \texttt{SS}& \texttt{SS} & \texttt{SS}
        \\
        \textbf{V}oltage& \texttt{0.8} & \texttt{0.8} &\texttt{0.9}&\texttt{0.9} &\texttt{0.88}&\texttt{0.88}&\texttt{0.88}&\texttt{0.99}&\texttt{0.99}&\texttt{0.99}&\texttt{0.72}&\texttt{0.72}&\texttt{0.72}&\texttt{0.81}&\texttt{0.81}&\texttt{0.81}
        \\
        \textbf{T}emperature& \texttt{25} & \texttt{85}& \texttt{25} & \texttt{85} &  \mbox{\texttt{-40}} & \texttt{0}& \texttt{125}& \mbox{\texttt{-40}} & \texttt{0}& \texttt{125}& \mbox{\texttt{-40}} & \texttt{0}& \texttt{125}& \mbox{\texttt{-40}} & \texttt{0}& \texttt{125}
        \\
        \multicolumn{17}{l}{
\multirow{2}{*}{\begin{tabular}[c]{@{}l@{}}
\\[-11pt]
$^\ast$ The units of voltage and temperature are V and $^\circ$C.\end{tabular}}
}\\[-9pt]
        \hline
        \end{tabular}
    \vspace{-6mm}
\end{table*}

%% file: fig/single.tex
\begin{figure}[]
\centering
\vspace{-0.2cm}
\captionsetup[subfigure]{justification=raggedright,singlelinecheck=false, margin={5pt,0pt}}
\begin{subfigure}{.24\textwidth}
  \centering
  \caption{DFF Setup vs. Delay}\label{fig:dff-delay}
  \vspace{-0.05cm}
  \includegraphics[width=4.4cm]{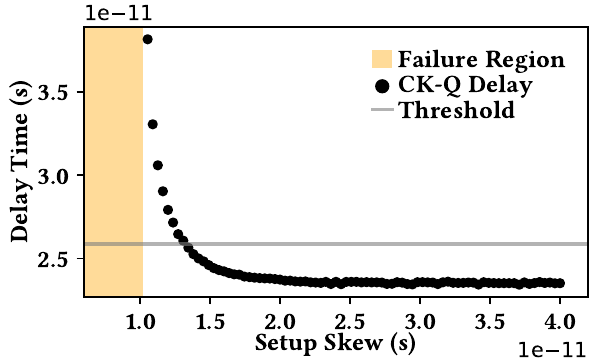}\\
  \vspace{-0.1cm}
  \caption{DFF Convergence Traces}\label{fig:dff-convergence}
  \vspace{-0.0cm}
  \includegraphics[width=4.4cm]{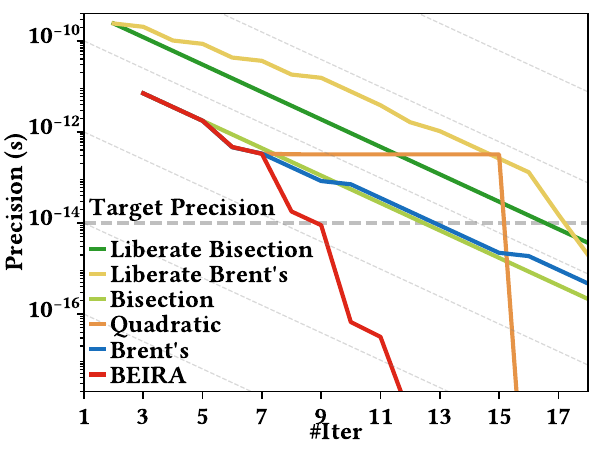}
\end{subfigure}%
\begin{subfigure}{.24\textwidth}
  \centering
  \caption{Latch Setup vs. Delay}\label{fig:latch-delay}
  \vspace{-0.05cm}
  \includegraphics[width=4.4cm]{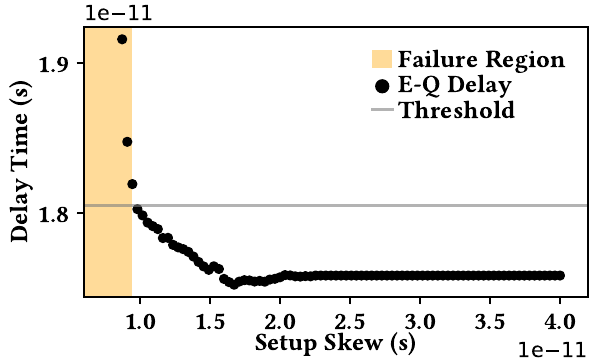}\\
  \vspace{-0.1cm}
  \caption{Latch Convergence Traces}\label{fig:latch-convergence}
  \vspace{-0.0cm}
  \includegraphics[width=4.4cm]{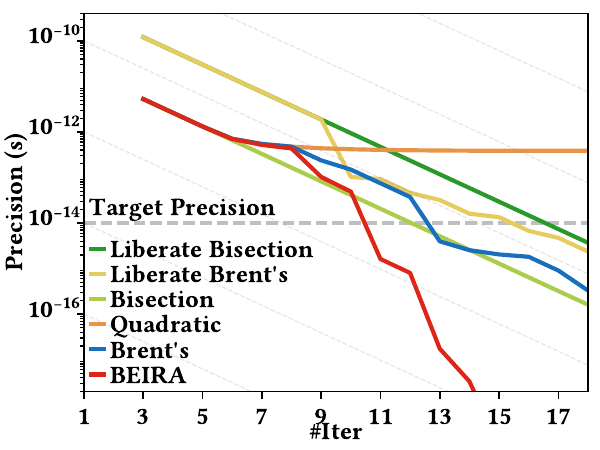}
\end{subfigure}
\vspace{-0.2cm}
\captionsetup{justification=centering}
\caption{Setup vs. Delay behavior for (a) DFF and (b) Latch; \\ All methods' convergence traces for (c) DFF and (d) Latch,\\the dotted reference lines refer to bisection's convergence.}
\label{fig:convergence}
\vspace{-0.5cm}
\end{figure}

%% file: fig/iteration.tex
\begin{figure*}[!t]
\vspace{-0.7cm}
\centering  
\begin{subfigure}{0.055\textwidth}
  \centering
  \parbox[c][3.6cm][c]{0.0507\textwidth}{%
    \centering \textbf{Init}%
  }\\[0.05cm]
  \parbox[c][3.6cm][c]{0.0507\textwidth}{%
    \centering \textbf{Iter 1}%
  }\\[0.05cm]
  \parbox[c][3.6cm][c]{0.0507\textwidth}{%
    \centering \textbf{Iter 4}%
  }\\[0.05cm]
  \parbox[c][3.6cm][c]{0.0507\textwidth}{%
    \centering \textbf{Sim Rest}%
  }\\
  \vspace{0.3cm}
\end{subfigure}
\begin{subfigure}{0.31\textwidth}
    \centering    
    \caption{Sample Selection}
    \label{fig:iter:selection}
    \includegraphics[height=3.6cm]{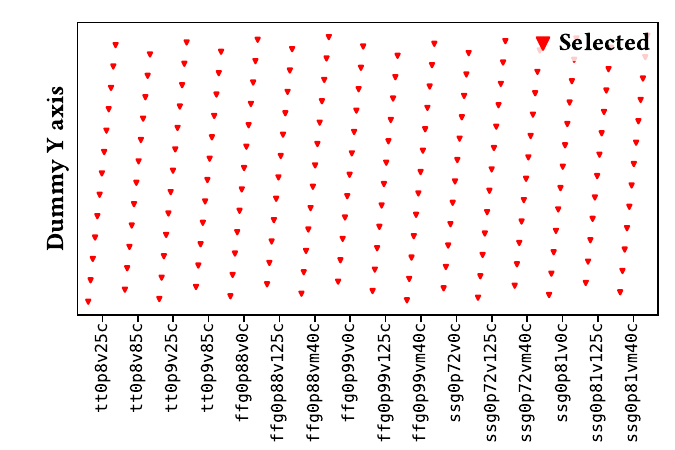}\\[0.00cm]
    \includegraphics[height=3.6cm]{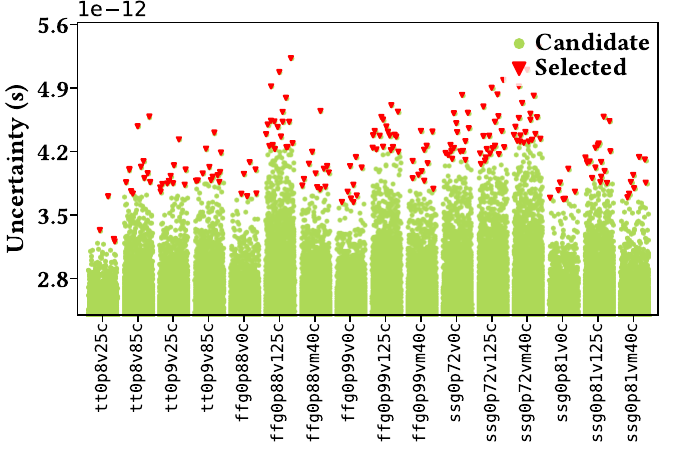}\\[0.00cm]
    \includegraphics[height=3.6cm]{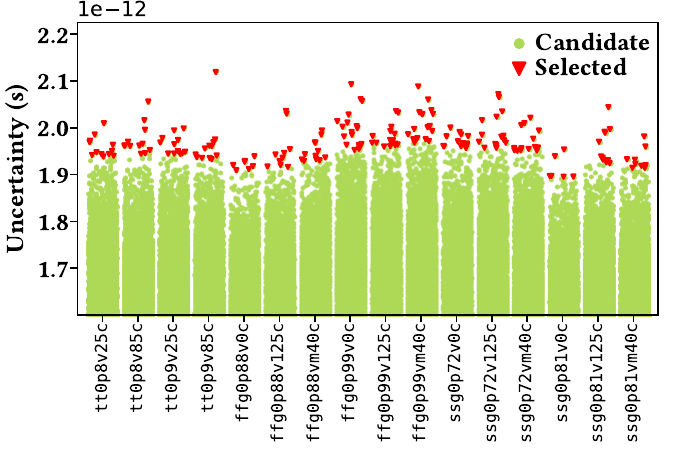}\\[0.00cm]
    \vspace{3.6cm}
\end{subfigure}%
\begin{subfigure}{0.31\textwidth}
    \centering
    \caption{Local Search Convergence}
    \label{fig:iter:simulation}
    \includegraphics[height=3.6cm]{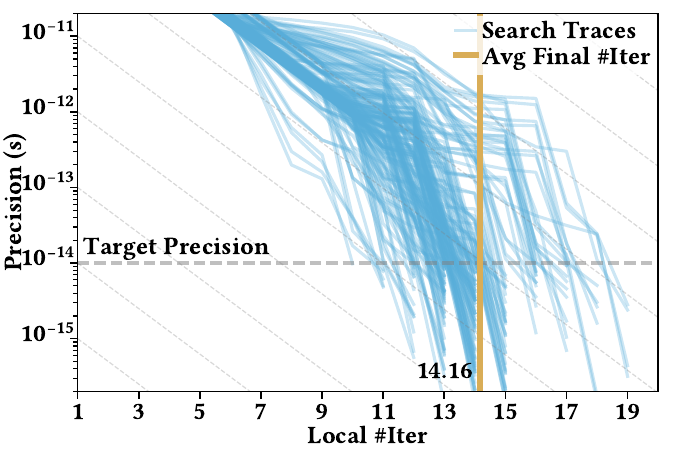}\\[0.00cm]
    \includegraphics[height=3.6cm]{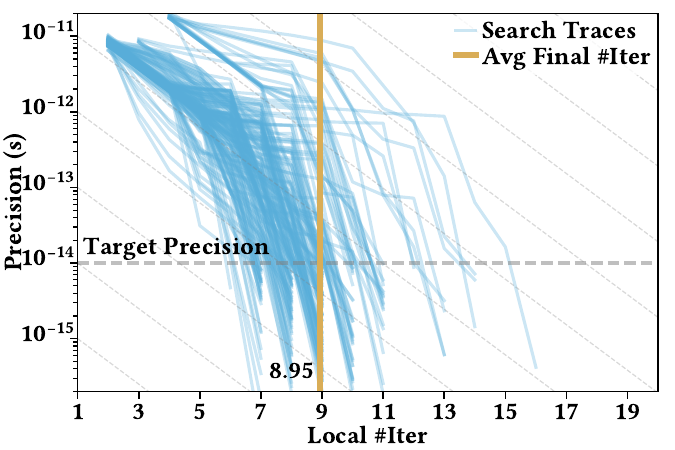}\\[0.00cm]
    \includegraphics[height=3.6cm]{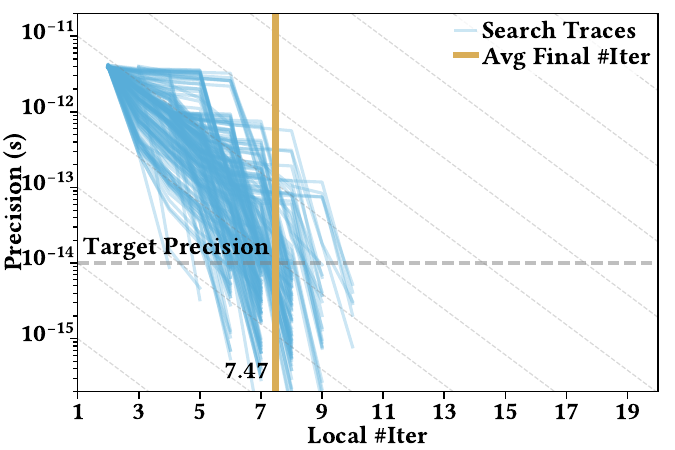}\\[0.00cm]
    \includegraphics[height=3.6cm]{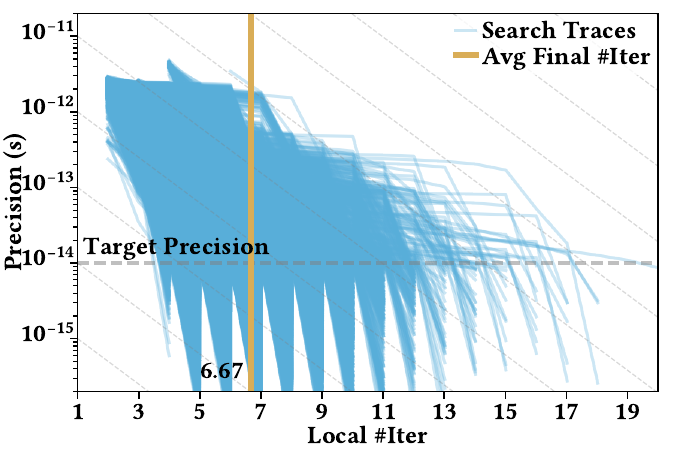}
\end{subfigure}%
\begin{subfigure}{0.23\textwidth}
    \centering
    \caption{Regression Fitness}
    \label{fig:iter:regression}
    \includegraphics[height=3.6cm]{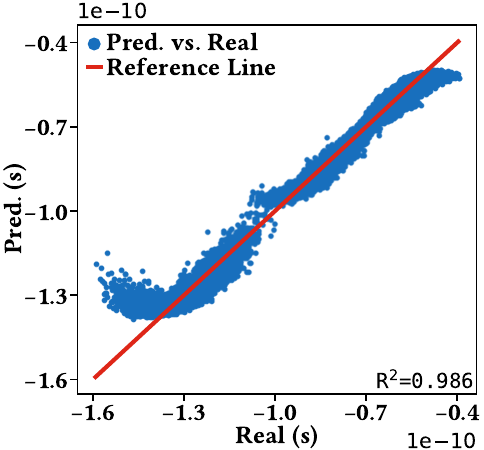}\\[0.00cm]
    \includegraphics[height=3.6cm]{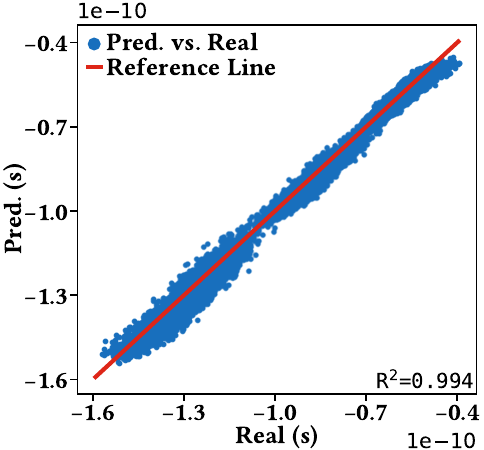}\\[0.00cm]
    \includegraphics[height=3.6cm]{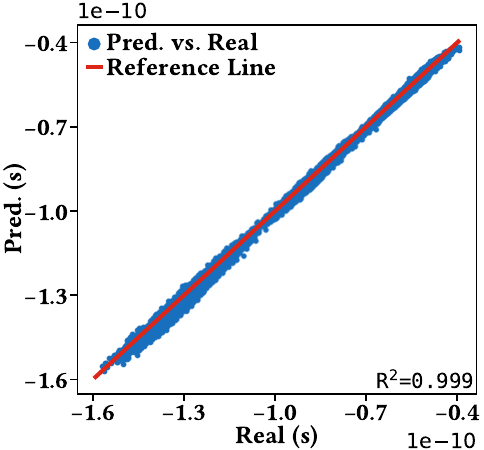}\\[0.00cm]
    \vspace{3.6cm}
\end{subfigure}%
\vspace{-.2cm}
\captionsetup{justification=centering}
\caption{(a) Candidate samples selection, (b) Trace of the search simulation (slanted dotted lines are reference lines for bisection convergence), and (c) Initial search interval for regression during the BEIRA active learning iteration. \\The initial interval from iteration 4 is then used to simulate the remaining samples.}\label{fig:iter}
\vspace{-0.7cm}
\end{figure*}

%% file: table/compare.tex

\begin{table}[!t]
\vspace{-0.1cm}
  \centering  	\renewcommand\arraystretch{0.98}
  \caption{Comparison of CPU Time$^\ast$ for one Setup/Hold Time Characterization.}\label{tab:compare}
  \vspace{-0.2cm}
\begin{tabular}{l|r@{\hspace{1pt}}r@{\hspace{22pt}}r@{\hspace{1pt}}r|r@{\hspace{1pt}}r@{\hspace{22pt}}r@{\hspace{1pt}}r}

\bottomrule
\\[-6pt]
\multirow{2}{*}{\textbf{\begin{tabular}[c]{@{}c@{}}
Procedure
\end{tabular}}} &  
\multicolumn{4}{c|}{\multirow{2}{*}{\begin{tabular}[c]{@{}c@{}} 
    \textbf{Bisection}\\ \ 
\end{tabular}}}&
\multicolumn{4}{c}{\multirow{2}{*}{\begin{tabular}[c]{@{}c@{}} 
    \textbf{BEIRA}\\ \ 
\end{tabular}}}
\\
\cmidrule(rl){2-5}
\cmidrule(rl){6-9}
\\[-14.2pt]
&&&&&&&&\\[-4pt]
&
\multicolumn{2}{l}{\hspace{2pt} basic} & 
\multicolumn{2}{c|}{w/ \textbf{AL}} & 
\multicolumn{2}{l}{\hspace{2pt} basic} & 
\multicolumn{2}{c}{w/ \textbf{AL}} \\
\hline
\\[-6pt]
SPICE&15.5&s&10.3&s&11.8&s&6.41&s\\
Search&1.91&ns&1.27&ns&186&µs&101&µs\\
AL&-&&17.0&ms&-&&16.9&ms\\

\toprule
\multirow{3}{*}{\textbf{\begin{tabular}[c]{@{}l@{}}
\\\\[2pt]
\hspace{-.2cm}\makebox[0pt][l]{\footnotesize $^\ast$  By averaging $25\text{ entry}\times16\text{ PVT}\times10k\text{ MC}=4M$ samples.}
\end{tabular}}}\\[-9pt]
Total&15.5&s&10.3&s&11.8&s&6.43&s\\
&
1&$\times$&
1.51&$\times$&
1.31&$\times$&
\textbf{2.41}&$\times$\\
\toprule
\end{tabular}
\vspace{-0.2cm}
\end{table}

%% file: reference.bib
@inproceedings{Naswali:2019,
  title={{DNNLibGen: Deep Neural Network Based Fast Library Generator}},
  author={Naswali, Eunice and Quiros, Adalberto Claudio and Chandran, Pravin},
  booktitle={2019 26th IEEE International Conference on Electronics, Circuits and Systems (ICECS)},
  pages={574--577},
  year={2019},
  organization={IEEE}
}

@inproceedings{Ma:2024,
  title={{Fast Cell Library Characterization for Design Technology Co-Optimization Based on Graph Neural Networks}},
  author={Ma, Tianliang and Deng, Zhihui and Sun, Xuguang and Shao, Leilai},
  booktitle={2024 29th Asia and South Pacific Design Automation Conference (ASP-DAC)},
  pages={472--477},
  year={2024},
  organization={IEEE}
}

@article{sobo:1967,
author = {I. M. Sobol},
title  = {{The Distribution of Points in a Cube
          and the Accurate Evaluation of Integrals}},
journal= {USSR Computational Mathematics and Mathematical Physics},
year   = 1967,
volume = 7,
number = 4,
pages  = {86--112}
}

@inproceedings{srivastava2007rapid,
  title={{Rapid and Accurate Latch Characterization via Direct Newton Solution of Setup/Hold Times}},
  author={Srivastava, Shweta and Roychowdhury, Jaijeet},
  booktitle={2007 Design, Automation \& Test in Europe Conference \& Exhibition},
  pages={1--6},
  year={2007},
  organization={IEEE}
}

@inproceedings{srivastava2007interdependent,
  title={{Interdependent Latch Setup/Hold Time Characterization via Euler-Newton Curve Tracing on State-transition Equations}},
  author={Srivastava, Shweta and Roychowdhury, Jaijeet},
  booktitle={Proceedings of the 44th annual Design Automation Conference},
  pages={136--141},
  year={2007}
}

@article{srivastava2008independent,
  title={{Independent and Interdependent Latch Setup/Hold Time Characterization via Newton--Raphson Solution and Euler Curve Tracking of State-transition Equations}},
  author={Srivastava, Shweta and Roychowdhury, Jaijeet},
  journal={IEEE Transactions on Computer-Aided Design of Integrated Circuits and Systems},
  volume={27},
  number={5},
  pages={817--830},
  year={2008},
  publisher={IEEE}
}

@article{guptacalibration,
  title={{Calibration of Setup and Hold Time for Latches and Flip-flops (II)}},
  author={Gupta, Puneet},
  publisher={Citeseer}
}

@manual{cadence2018liberate,
  title        = {{Virtuoso Liberate Reference Manual}},
  organization = {Cadence Design Systems, Inc.},
  edition      = {Product Version 17.1},
  month        = {March},
  year         = {2018},
}

@manual{synopsys2024primelib,
  title        = {{PrimeLib User Guide}},
  organization = {Synopsys, Inc.},
  edition      = {Version W-2024.09},
  month        = {September},
  year         = {2024},
}

@book{brent2013algorithms,
  title={{Algorithms for Minimization without Derivatives}},
  author={Brent, Richard P},
  year={2013},
  publisher={Courier Corporation}
}

@inproceedings{zhou2025lvfgen,
  title={{LVFGen: Efficient Liberty Variation Format (LVF) Generation Using Variational Analysis and Active Learning}},
  author={Zhou, Junzhuo and Lin, Ting-Jung and Xia, Haoxuan and Huang, Li and Xing, Wei and He, Lei},
  booktitle={Proceedings of the 2025 International Symposium on Physical Design},
  pages={182--190},
  year={2025}
}

@book{Sharma:2015,
author = {Sharma, Rohit},
year = {2015},
month = {11},
pages = {},
title = {{Characterization and Modeling of Digital Circuits}},
isbn = {1500733679}
}

@inproceedings{roethig2003library,
  title={{Library Characterization and Modeling for 130 nm and 90 nm SOC Design}},
  author={Roethig, Wolfgang},
  booktitle={IEEE International [Systems-on-Chip] SOC Conference, 2003. Proceedings.},
  pages={383--386},
  year={2003},
  organization={IEEE}
}

@inproceedings{phelps1991advanced,
  title={{Advanced Library Characterization for High-Performance ASIC}},
  author={Phelps, RW},
  booktitle={Proceedings Fourth Annual IEEE International ASIC Conference and Exhibit},
  pages={P15--3},
  year={1991},
  organization={IEEE}
}

@INPROCEEDINGS{Onaissi2011,
  author={Onaissi, Sari and Taraporevala, Feroze and Liu, Jinfeng and Najm, Farid},
  booktitle={2011 48th ACM/EDAC/IEEE Design Automation Conference (DAC)}, 
  title={A fast approach for static timing analysis covering all PVT corners}, 
  year={2011},
  volume={},
  number={},
  pages={777-782},
  doi={}}
